\documentclass[fleqn,usenatbib]{mnras}

\usepackage{newtxtext,newtxmath}

\usepackage[T1]{fontenc}
\usepackage{ae,aecompl}

\usepackage{graphicx}	
\usepackage{amsmath}	
\usepackage{multicol}
\usepackage[normalem]{ulem}

\usepackage[dvipsnames]{xcolor}

\title[Molecular Gas Initial Conditions]{Using Molecular Gas Observations to Guide Initial Conditions for Star Cluster Simulations}

\author[A. Sills et al.]{
Alison Sills,$^{1}$\thanks{E-mail: asills@mcmaster.ca}
Steven Rieder,$^{2,3}$\thanks{E-mail: steven.rieder@unige.ch}
Anne S.M. Buckner,$^{3}$
Alvaro Hacar,$^{4}$
\newauthor
\,\,Simon Portegies Zwart,$^{5}$
and Paula S. Teixeira,$^{6}$
\\
$^{1}$ Department of Physics \& Astronomy, McMaster University, 1280 Main Street West, Hamilton ON, L8S 4M1, Canada\\
$^{2}$ Geneva Observatory, University of Geneva, Chemin Pegasi 51, 1290 Sauverny, Switzerland\\
$^{3}$ School of Physics and Astronomy, University of Exeter, Stocker Road, Exeter, EX4 4QL, UK\\
$^{4}$ Department of Astrophysics, University of Vienna, T\"urkenschanzstrasse 17, 1180 Vienna, Austria\\
$^{5}$ Leiden Observatory, Leiden University, NL-2300RA Leiden, the Netherlands\\
$^{6}$ SUPA, School of Physics and Astronomy, University of St. Andrews, North Haugh, St. Andrews KY16 9SS, Fife, UK\\
}

\date{Accepted 2022 December 09. Received 2022 December 01; in original form 2022 September 16}

\pubyear{2022}

\begin{document}
\label{firstpage}
\pagerange{\pageref{firstpage}--\pageref{lastpage}}
\maketitle

\begin{abstract}
The earliest evolution of star clusters involves a phase of co-existence of both newly-formed stars, and the gas from which they are forming. Observations of the gas in such regions provide a wealth of data that can inform the simulations which are needed to follow the evolution of such objects forward in time. We present a method for transforming the observed gas properties into initial conditions for simulations that include gas, stars, and ongoing star formation. We demonstrate our technique using the Orion Nebula Cluster. Since the observations cannot provide all the necessary information for our simulations, we make choices for the missing data and assess the impact of those choices. We find that the results are insensitive to the adopted choices of the gas velocity in the plane of the sky. The properties of the surrounding gas cloud (e.g. overall density and size), however, have an effect on the star formation rate and pace of assembly of the resultant star cluster. We also analyze the stellar properties of the cluster and find that the stars become more tightly clustered and in a stronger radial distribution even as new stars form in the filament. 
\end{abstract}




\section{Introduction}

Understanding the formation and earliest evolution of star clusters involves the physics of the interstellar medium, stellar dynamics, stellar evolution, and the interplay between all three. Stars form over a period of a few million years, in close proximity of each other inside giant molecular clouds. There is a phase during which the evolving stars are exerting an influence on the gas through their radiation, winds, and eventually supernovae, all of which act to remove the gas from the cluster. Molecular clouds are turbulent, and the resulting initial stellar distribution is usually clumpy and quite sub-structured\citep[e.g.][]{McKee2007}. Any stars that make up a bound structure will relax into a spherical system fairly quickly. The crowded environment in which stars form and spend their early lives could affect stellar properties such as the distribution of binary parameters \citep{CCC2021}, triples and higher order multiples \citep{2007MNRAS.379..111V}, and the properties of planetary systems \citep{Parker2009}. We want to understand the processes that convert a cloud of gas into a spherical, bound, gas-free star cluster or a field of unbound stars, or both. The timescales for these processes, and the conditions which determine the fraction of the resultant stars in clusters, are important for understanding stellar populations in galaxies. 

One approach to answering these questions is observational. Detailed photometric studies of young local star-forming regions in the optical, infrared, and X-ray have been used to constrain the spatial distributions of young stellar objects \citep[e.g.][]{Spitzer,Kuhn2015}, and the addition of Gaia data has allowed for wider explorations in the velocity realm to look at the expansion or contraction of the clusters \citep[e.g.][]{Kuhn2019}. Timescales are inferred by comparing different regions and trying to use pre-main sequence isochrone fitting to date the systems \citep[e.g.][]{DaRio2010}. One limitation of using only the stars to understand early cluster properties is that at the earliest, deeply embedded stages, the gas is the dominant component of the region, and the stars are difficult to observe, especially in the optical and infrared.
 
Even after much of the gas has been removed from the region, extinction can still be large, and therefore the (optical) Gaia coverage of such regions is poor, which reduces the number of stars for which we have reasonable proper motion measurements. Radial velocities are possible to obtain from infrared spectra \citep[e.g.][]{Cottaar2015}, but obtaining spectra for many stars in a cluster is an expensive prospect with current instruments. Age, mass, and distance measurements of the stars are also difficult, because of extinction and the necessity of working in the infrared, but also because pre-main sequence stars are variable which affects their position in the HR diagram \citep{Messina2017}.

Stellar observations also ignore the other major component in these forming clusters, namely the gas. Surveys of local star-forming regions in the far infrared, such as those from {\it Herschel}, use continuum dust emission to map out the column density distribution of material \citep[e.g.][]{Lombardi2014}. This provides an estimate of the mass in the cloud. Molecular line observations in the (sub-)millimeter provide velocity information along the line of sight, temperature information, and column density \citep[e.g.][]{Hacar2017}. And because different lines trace different density ranges of gas, a combination of lines can nicely map out the gas distribution in the region. Unfortunately, there is almost no possibility of obtaining information about gas proper motions, and the depth of gas clouds can only be inferred in specific cases using density estimates \citep[][]{Li2012}.

A complementary approach to answering questions about the earliest stages of star cluster formation is to use computational approaches. Simulations allow us to probe in great detail all the physical processes that are going on, and we can also follow the systems in three dimensions and in time to understand how they will evolve. However, the initial conditions for these kinds of simulations are almost always idealistic. Gas and star distributions are often spherical, usually in virial equilibrium, and with a density distribution that is straightforward to set up (e.g. constant density for gas although see \citet{Chen2021} for an exploration of more realistic distributions, and a Plummer or King model for stars). Simulations have traditionally been optimized for only one of the components -- hydrodynamics for gas; N-body dynamics for stars -- and an approximation has been included for the other. Hydrodynamics simulations often use sink particles to represent stars or regions which will form stars; N-body codes can include an analytic background potential to mimic the presence of gas. An increasing number of groups are using different techniques to improve the realism of simulations, such as including both stellar dynamics and hydrodynamics simultaneously \citep[e.g.][]{wall2019}, or using the results of hydrodynamics simulations as initial conditions for N-body runs \citep[e.g.][]{Fujii2016,Ballone2021}. In order to be most useful, simulations must include all the appropriate physics for the question at hand, and of course they must start with appropriate initial conditions. 

Our hybrid approach is to bridge the divide between observations and simulations, and investigate the impact of more realistic, observationally-motivated initial conditions on our understanding of the formation and evolution of star clusters. In a previous paper \citep{Sills2018} we used observed positions of stars in embedded young clusters as a starting point for stellar dynamics plus hydrodynamics simulations. We showed that the initially clumpy sub-structure of the clusters evolved towards a more spherical distribution over a relatively short time. In those simulations, we included the presence of gas in a somewhat idealized way (following the clumpy distribution of the stars) and we showed that the total amount of gas present could have a significant impact on the subsequent evolution of the cluster. 

In this paper, we take the next step to improve the realism of this kind of simulation. In addition to the observed properties of stars, we include observed properties of the dense molecular gas in simulations of young embedded star clusters. Our goal is to explore realistic conditions for cloud evolution, as we expect that the idealized simulation environment that has been used in previous work may not correctly capture the impact of a range of gas densities and morphologies on the star formation rate and dynamical evolution of the forming star cluster. Specifically we look at the Orion region, near the Orion Nebula Cluster and its Integral Shaped Filament \citep[ISF; ][]{JOH99}. We describe the methods by which we use the observational data to dictate the properties of our simulated gas, and highlight some tests we used to determine which choices of non-observed parameters were the most reasonable. Finally, we predict some likely outcomes of the star formation and dynamical evolution of the Orion region.

\section{Methods}

\subsection{Computational Framework}

For the simulations in this paper, we use the {\tt Ekster} code \citep[][]{Rieder2022, ekster}.
This code combines the SPH code {\tt Phantom} \citep{Phantom}, the $N$-body code {\tt PeTar} \citep{Petar} and the stellar evolution code {\tt SeBa} \citep{Seba1, Seba2} within the {\tt AMUSE} framework \citep{AmuseBook,amuse2013a,amuse2013b,amuse2009}.

The main benefit of {\tt Ekster} is that it allows for the formation of individual stars in a simulation that would otherwise not have the required resolution to form these directly from the gas.
We restrict ourselves to an isothermal EOS in this study, and feedback effects (like stellar winds and ionising radiation) are not taken into account. Star formation in {\tt Ekster} works similarly to the method described in \citet{wall2019}.
Once gas reaches a density $\rho_{\rm crit}$ (here: $10^{-18}$ g cm$^{-3}$ or $~ 1.5\times10^4$ M$_{\odot}$ pc$^{-3}$), it will form a sink particle if the gas is collapsing \citep[see][for more specific criteria]{Rieder2022}.
This sink will then accrete gas within a radius of 0.1 pc.
A sink is considered as a source of star formation, and will form individual stars by probing a \citet{Kroupa2001} Initial Mass Function (IMF) until the next star is more massive than the remaining mass in the sink.
We use the modified `grouping' method described in \citet{Liow2022}, in which groups of sinks act as a combined star-forming region.
This has the benefit over the original `single sink' method of not requiring a high mass for each sink in order to probe the IMF and still form massive stars.

\subsection{Initial Conditions}

In order to run a simulation, we need (at a minimum) 7 parameters for stars and 8 parameters for gas.
For stars, we must specify their mass and 3-D positions and velocities; for gas particles, we need those same parameters and also a temperature or internal energy.
However, the observations are incomplete in both cases -- typically we can determine quantities only in the plane of the sky (e.g. positions from imaging, or velocities from proper motions), or only along the line of sight (e.g. velocities from spectra, or distances from parallax).
In this section we describe which quantities we take directly from observations of the Orion region, and the choices we make for the other quantities. We note that the gas particles could be given additional properties, such as a chemical composition, dust-to-gas ratio, or a magnetic field. When such quantities are available from observations, they should also be included wherever possible. In this suite of simulations, we assume a constant mean molecular weight of 2.4 for our gas particles, no dust, and we neglect magnetic fields.

\subsubsection{Stars}
The initial properties of the stars are chosen as described in \citet{Sills2018}, with some additions to extend the stellar distribution to larger distances.
Here, we briefly summarize our method.
We use stars from the Massive Young Stars in the Infrared and Xray (MYStIX) \citep{Feigelson2013} survey, and their clustering properties as described in \citet{Kuhn2015}.
We use the observed positions of these stars (right ascension and declination) and take the distance to Orion to be 414 pc \citep{Menten2007}.
This determines the stellar position in the x and y plane of our simulation.
Most of these objects are not detected by Gaia, so the line of sight (z) position of each star is drawn randomly between -0.5 and 0.5 pc.
The masses of the observed stars are drawn randomly from a \citet{Kroupa2001} IMF between 0.8 and 10 $M_{\odot}$, corresponding to the observational limits of the survey.
An appropriate number of lower mass stars (down to 0.2 $M_{\odot}$) were added to complete the IMF, and were distributed randomly within Plummer ellipsoids whose sizes on the sky were determined by the size and orientation of the MYStIX subclusters.
We also included a sphere of lower mass stars to complete the IMF down to 0.2 $M_{\odot}$ corresponding to the `unclustered/unknown' component as identified in MYStIX. 

The MYStIX region only covers the central portion (r $\approx$ 1 pc) of the area covered by the gas observations.
We used stars identified by Spitzer \citep{Spitzer} in an annulus between 1 and 5 pc from the centre of the region to complete this region.
We treated these stars in the same way as the MYStIX stars: identified Spitzer sources were assigned masses between 0.8 and 10 $M_{\odot}$, and their line-of-sight position was randomly selected in the same range as the MYStIX stars.
We then populate the IMF with the correct number (according to the adopted IMF) of lower mass stars, and randomly distributed them in this spherical shell.
Finally, the Orion Cluster has 6 stars with masses greater than 10 $M_{\odot}$. We used the observed positions and masses of those stars from \citet{Hillenbrand1997}, and assigned them z positions randomly distributed within 0.025 pc of z=0.2 (i.e. slightly in front of the gas, as suggested by \citet{Wen1995}). We note that we are explicitly neglecting the binarity/multiplicity properties of stars in these simulations and assuming that all stars are single. The presence of binaries will modify the stellar dynamics in these simulations, but we do not expect them to have a large effect on the gas.
All stars were then given a random velocity such that the overall velocity dispersion was 1 km s$^{-1}$. 

\subsubsection{Gas}

To describe the initial gas conditions we use molecular line observations along the Integral Shape Filament in Orion. 
We use N$_2$H$^+$ (1-0) observations tracing gas with densities higher than n(H$_2$) = $10^{4.5}$ cm$^{-3}$ together with ancillary NH$_3$-based gas temperatures estimates in this cloud taken from \citet{Hacar2017} and \citep{Friesen2017}, respectively, both Nyquist sampled at 30 arcsec resolution. There is also lower-density gas in the Orion region which we will include, but with properties not taken directly from observations.

This combined dataset provides the column density, mass, temperature, line of sight velocity, and line of sight velocity dispersion for dense gas in pixels distributed along the filament.
Each pixel is 0.03 pc (equivalent to 15 arcsec sampling) on a side at the distance of Orion. We wish to distribute gas particles in our simulation volume so that they have the same properties as the observations.
We choose a mass for our gas particles (0.02 $M_{\odot}$ in these simulations) and then determine how many such particles are needed to match the mass in each pixel.
If a pixel contains less mass than our particle mass, we do not place a particle in the simulation.
If the pixel contains enough mass for 1 particle then it is placed at the centre of the pixel in the z=0 plane.
If more than one gas particle is required, they are placed randomly within the pixel (uniformly sampling in x and y), and the particles are given a random position in z between 0.15 pc and -0.15 pc.
The total mass in dense gas is 1225 M$_{\odot}$.
The temperature of each particle is determined by the measured temperature in that pixel, 
which range between 10~K in the densest parts of the cloud away from the Trapezium and 137~K in the vicinity of the Orion BN/KL region. This temperature is used to determine the internal energy of the particle, assuming a mean molecular weight of 2.4.

The velocities of these parcels of gas along the line of sight are determined by the observations.
We remove a systematic mean motion of the observed gas (9.74 km~s$^{-1}$ away from us) and assign each particle the resultant velocity within its pixel along the z direction of our simulation.
If there is more than one particle within a pixel, we modify their velocities so that they are drawn from a Gaussian distribution with a mean given by the gas velocity relative to the local standard of rest of the pixel, and a dispersion given by the velocity dispersion as measured in that pixel. 

We have no observed information about the velocity of the gas in the plane of the sky.
Therefore, we need to make some choices in order to have a reasonable representation of the gas properties.
Those choices, and the tests we performed to assess which ones are most reasonable, are described in more detail in the following section. 

The dense gas observations do not capture the entire interstellar medium in this region.
The whole area is filled with relatively cool, low density gas as well.
We add a sphere of gas, with a temperature of 30~K, a density of 500 particles per cm$^3$, and a radius of 5 pc to our simulation.
This results in a total of $1.57 \times 10^4$ M$_{\odot}$ of low-density gas. 
Specifically, we used the \texttt{molecular\_cloud} routine within the \textsc{AMUSE} framework \citep{AmuseBook,fpz15} which includes a turbulent power spectrum of the form $k^{-4}$ for the gas velocities.
Our virial ratio, defined as the ratio of kinetic to potential energy, was set to 1.0, leading to a 3D velocity dispersion of about 4 km s${^-1}$. We also varied many of these parameters of this gas sphere, and we will discuss the effect of other choices on our simulations below.

\begin{figure*}
    \centering
	\includegraphics[width=\textwidth]{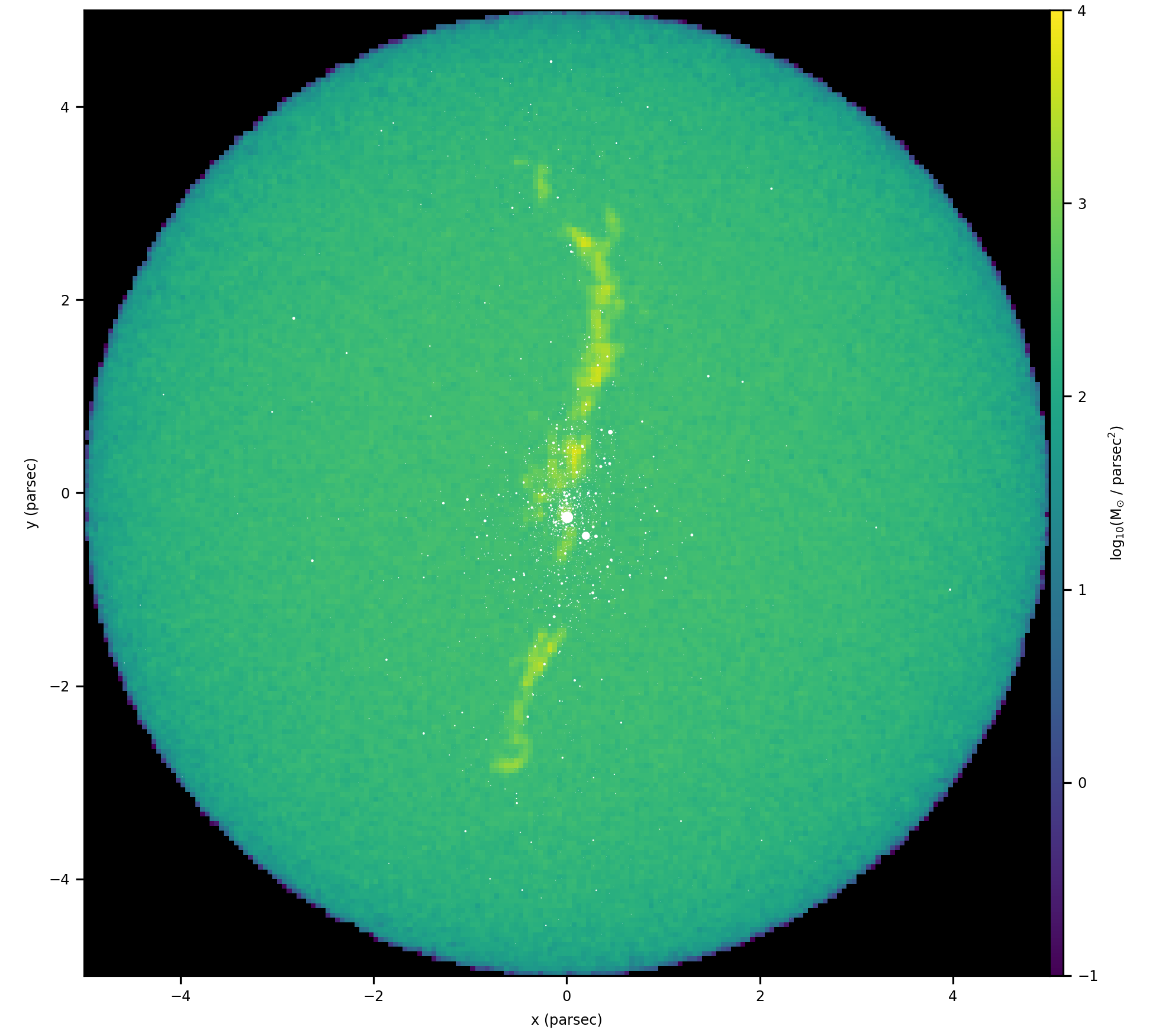}
    \caption{Projected column density of gas in the xy plane (plane of the sky) is shown as the colour scale and the projected positions of the stars are shown as white dots, with the size of the dot directly proportional to stellar luminosity. The most massive stars/largest dots are about 30 M$_{\odot}$.}
    \label{fig:xy}
\end{figure*}

Figure \ref{fig:xy} shows an image of the xy plane of our default choice of initial conditions (model vx\_vy\_0) at t=0.
The colour scheme gives the column density of gas and the stars are plotted as white circles, with the circle size proportional to stellar mass.
In all our simulations, the placement and properties of the stars are the same.
The 1225 $M_{\odot}$ of observed dense gas is modelled with almost 60 000 particles.
In this case, the background sphere of gas is at a constant density of 30 $M_{\odot} pc^{-3}$ and has a radius of 5 pc.
As expected, the distribution of gas in the plane of the sky looks like a realistic filament of gas near a star cluster. 

\begin{figure}
    \centering
	\includegraphics[width=\columnwidth]{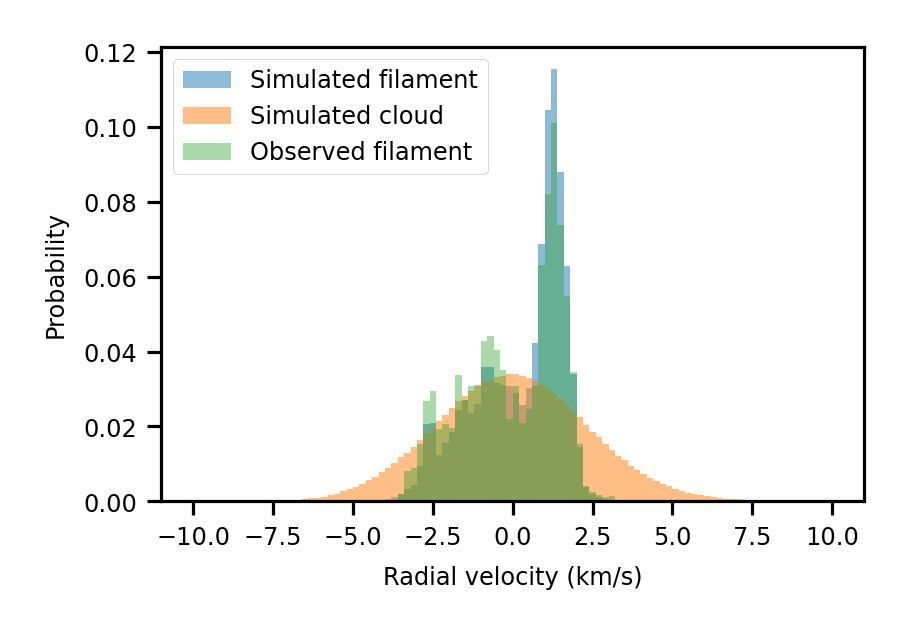}
    \caption{ Histograms of particle or pixel velocities for gas along line of sight (z axis). The green histogram gives the observed velocities for the dense gas, and those are well-matched with the blue histogram of the particles that model that component. The orange histogram shows the velocities of particles in the low-density turbulent sphere with a virial ratio of 1.0, which are centred at zero by construction.}
    \label{fig:velocity}
\end{figure}

Figure \ref{fig:velocity} shows the distribution of line-of-sight (z) velocities for particles in our dense gas filament (blue) and sphere of low-density gas (orange) compared to the observed line-of-sight velocities in the filament (green).
The particles in the filament follow the observations closely, confirming that our initialization of those particles is correct to the extent constrained by the data. The particles in the sphere have velocities set by a turbulent power spectrum and, in this particular case, have a virial ratio of 1.0.
The observed velocity structure of the filament is well-reproduced by our method, and shows more structure than the artificially-created low-density background cloud. That cloud does have a reasonable velocity dispersion, centred at zero by construction. 

\begin{figure}
    \centering
    \includegraphics[width=\columnwidth]{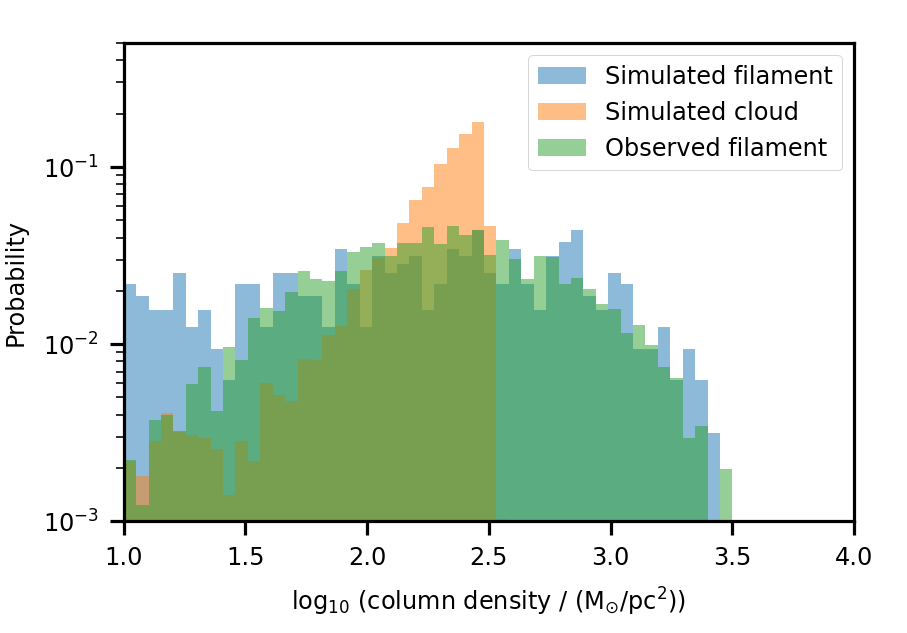}
    \caption{Probability density function of the gas column density at t=0 in the simulation shown in Figure \ref{fig:xy}. The blue histogram shows the observed dense filament, the orange histogram show the constant density sphere of background gas, and the green histogram shows the column density PDF of the observations. The simulated filament matches the observations at the high density end; the excess of gas at low densities shows the effect of having a minimum gas particle mass of 0.02 $M_{\odot}$.
    }
    \label{fig:PDF}
\end{figure}

In Figure \ref{fig:PDF} we show the probability density function (PDF) of the column density of the gas at t=0.
This is a common quantity calculated for observed molecular clouds and star-forming regions \citep[e.g.][]{Lombardi2014}.
In observations, the column density PDF is often seen to take a log-normal shape, sometimes with a power law tail to high column densities \citep{Kainulainen2009}. 

The blue histogram is the column density of the filament alone, and the orange histogram shows the column densities of the background cloud.
Our simulated gas is the sum of these two.
The green histogram shows the observed column density distribution, and at 100 $M_{\odot} pc^{-2}$ and higher, the simulation correctly represents the observations. 
The excess of power in the simulations compared to the observations at lower densities is a product of our choice of a gas particle mass of 0.02 $M_{\odot}$.

\section{Results}

\subsection{Evolution of system with the default parameters}

\begin{figure}
\centering 
\includegraphics[width=\columnwidth]{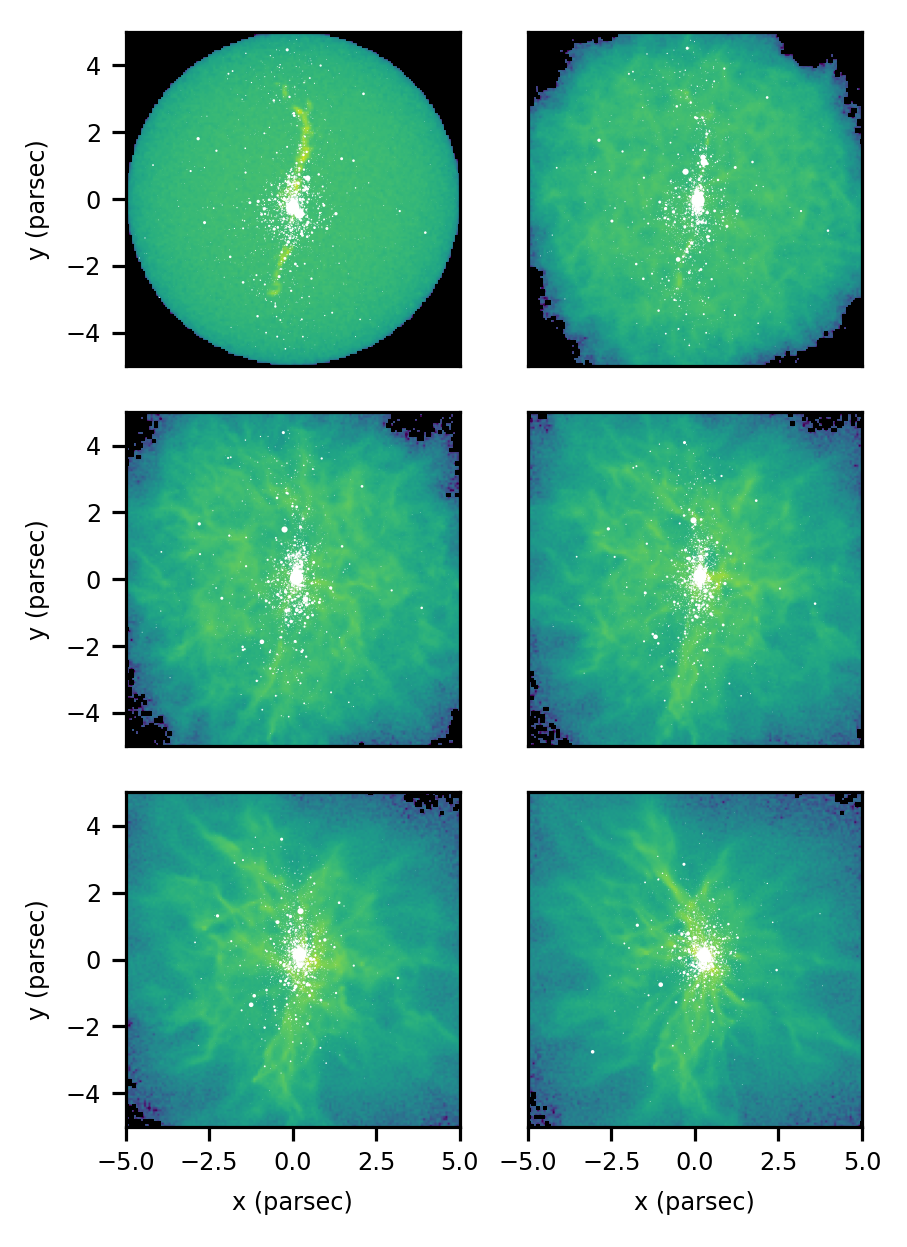}\par
    \caption{Evolution of model vx\_vy\_0. Each panel shows the projected column density and stellar positions as in Figure \ref{fig:xy}. The timestep from the top left to bottom right is 0.2 Myr. The collapse of the background cloud is very obvious. The gas along the dense filament collapses along the filament width and quickly flows into the central star cluster, forming new stars as it does.
    }
    \label{fig:evolution_vxy0}
\end{figure}

In Figure \ref{fig:evolution_vxy0} we show 1 Myr of evolution of our simulation with the initial conditions described above and shown in the previous figures.
Very quickly, the dense gas in the filament collapses towards the centre ridge line of the filament, and stars begin to form.
The filament gas, and also the recently formed stars, fall into the central star cluster.
At the same time, the larger sphere of gas develops a density structure driven by the initial turbulent velocity spectrum.
As some of that gas reaches high enough density, stars begin to form further away from the filament.
Finally, the bulk of the low-density gas also falls into the central region of the simulation, and the stellar system becomes more spherical. 

\begin{figure}
    \centering
    \includegraphics[width=\columnwidth]{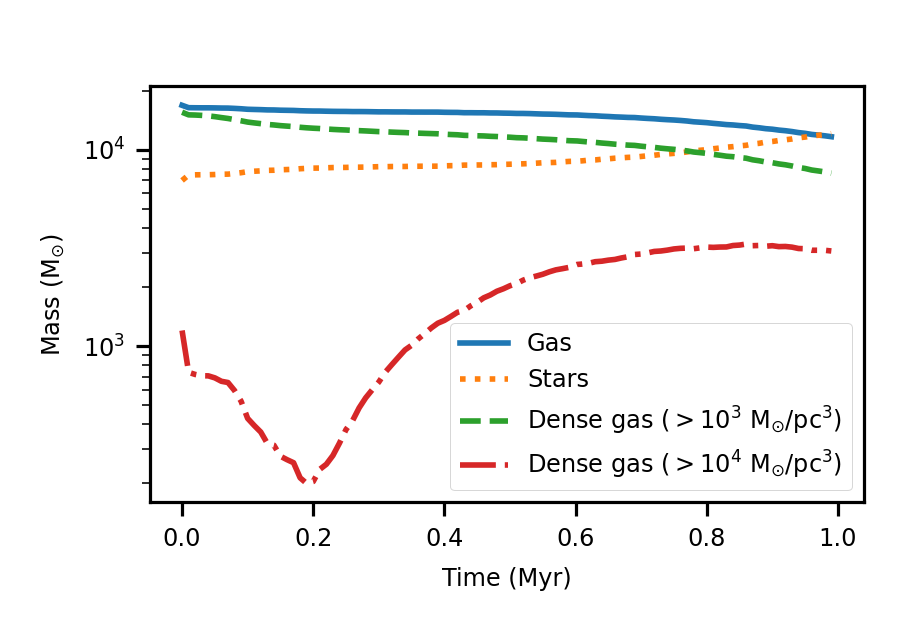}
    \caption{Masses of components of the simulation as a function of time for model vx\_vy\_0. The blue solid and orange dotted lines show the total mass in gas and stars, whereas the other lines show the mass of dense, star-forming gas only -- above $10^3$ M$_{\odot}$pc$^{-3}$ for the green dashed line and above $10^4$ M$_{\odot}$pc$^{-3}$ for the red dot-dashed line.
    }
    \label{fig:masses}
\end{figure}

Figure \ref{fig:masses} shows the mass of various simulation components as a function of time.
Over the 1 Myr of this simulation, the gas is converted to stars to the point where the two contributions are approximately equal.
The dense gas (all gas above $10^3 M_{\odot}pc^{-3}$) decreases slightly faster than the total gas mass.
There is always a background of low-density gas that does not participate in star formation, which becomes a larger fraction of the gas mass as more and more stars are formed.
We can see that the densest gas (above $10^4 M_{\odot}pc^{-3}$, originally just in the filament) is quickly depleted (within the first 0.2 Myr) as the filament collapses, and the star formation rate peaks at a very high rate of 4000 $M_{\odot}$ Myr$^{-1}$.
After a short pause, the background cloud's collapse replenishes this reservoir of dense gas, and in fact increases the amount available for star formation.
By about 0.5 Myr, the amount of densest gas begins to level off, but the large amount available means that the star formation rate at this point in the cluster's evolution is high and increasing, becoming larger than 10 000 $M_{\odot}$ Myr$^{-1}$.
Estimates of the average star formation rate in the solar neighbourhood are around 2500 $M_{\odot}$ Myr$^{-1}$ based on studies of star clusters \citep{Bonatto2011}, and less than 1000 $M_{\odot}$ Myr$^{-1}$ in local star-forming clouds \citep{Lada2010}.
Our star formation rates are far too high, and the timescale of collapse of the filament (about 0.2 Myr or so) is extremely quick.
In this particular model, the filament has no velocity in the plane of the sky so this collapse is not unexpected.
We will explore other choices for that velocity in the next section. 

\begin{table*}
\centering
\caption{Initial conditions for simulations}
\label{table:models}
\begin{tabular}{lcccl}
Model Name     & net xy velocity & size of sphere & Virial Ratio & Density of Spherical Cloud \\
\hline
vx\_vy\_0     & 0 km s$^{-1}$ & 5 pc & 1.0 & 30 M$_{\odot}$ pc$^{-3}$\\
vx\_vy\_vz & z velocity & 5 pc & 1.0 & 30 M$_{\odot}$ pc$^{-3}$\\
vx\_vy\_grad & gradient & 5 pc & 1.0 & 30 M$_{\odot}$ pc$^{-3}$\\
vx\_vy\_grad\_2.0 & gradient & 5 pc & 2.0 & 30 M$_{\odot}$ pc$^{-3}$\\
vx\_vy\_grad\_10pc & gradient & 10 pc & 1.0 & 30  M$_{\odot}$ pc$^{-3}$\\ 
vx\_vy\_grad\_10pc\_3.75 & gradient & 10 pc & 1.0 & 3.75  M$_{\odot}$ pc$^{-3}$\\ 
Idealized & none & 10 pc & 1.0 & 32.3 M$_{\odot}$ pc$^{-3}$\\
\hline
\end{tabular}
\end{table*}

\subsection{Free parameters in gas initial conditions}

In this section, we describe in more detail the choices we made to fill in the missing information for the observed dense gas and for the low-density gas background.
Each of the model parameters are summarized in Table \ref{table:models}.
We changed two general properties -- the velocity of the dense gas in the plane of the sky, and the parameters of the background sphere of gas. 


We have velocity information for the dense gas along the line of sight, but no information in the plane of the sky (the xy plane of our simulation).
Therefore, we have tested three different possibilities, as described below.

\begin{enumerate}
        \item In model vx\_vy\_0 we set the bulk x and y velocities of each particle to 0 km s$^{-1}$. This is clearly not physical but provides a lower limit to the total amount of kinetic energy from the bulk velocity of the dense gas.
        \item In model vx\_vy\_vz we set the net x and y velocities to the same value as the net z velocity in each pixel. This assumes that the filament is symmetric along all three axes. 
        \item In model vx\_vy\_grad, we use the gradient in line-of-sight velocities as a proxy for the velocities in the plane of the sky. For each pixel, the net line of sight velocity was compared to the line of sight velocity in the pixel directly above, and directly below (in declination, or the y direction), the pixel in question.  The net y velocity for that pixel was taken to be the difference in line of sight velocities between those two pixels (i.e. the local gradient), and we preserved the direction of the gradient as well. If there was only information in one of the two pixels (either above, or below, but not both) then the gradient was calculated using the velocity of the pixel that we are considering and the one pixel above/below. If there was no information either above or below the pixel, the net y velocity was taken to be 0 km s$^{-1}$. Similarly, the x velocity was taken using the gradient of the line-of-sight velocity in the pixels to the left and right of the pixel in question (i.e. in the right ascension direction).
\end{enumerate}


Our default simulation sets the velocity dispersion in the background gas assuming it is in virial equilibrium (with itself only -- not including the stars and the dense gas), i.e. with a ratio of kinetic to potential energy of 1.0 as in \citet{fpz15}.
We also ran a super-virial simulation (vx\_vy\_grad\_2.0), with a ratio of kinetic to potential energy of 2.0.
Finally, we also increased the size of the sphere from 5 pc to 10 pc.
In the first case (vx\_vy\_grad\_10pc) we kept the density of the gas the same but added more to the cloud (i.e. increased the mass by a factor of 8) and in a second case (vx\_vy\_grad\_10pc\_3.75) we kept the mass the same (i.e. reduced the density by a factor of 8). 

For comparison purposes, we have also set up a simulation in which the gas is initialized with a distribution more like the traditionally, idealized initial conditions for gas simulations.
We removed the filament, and added the total amount of mass in the dense gas observations to a spherical cloud.
We set the radius of this cloud to 10 pc, and gave it a virial ratio of 1.0. 

\begin{figure}
    \centering
        \includegraphics[width=\columnwidth]{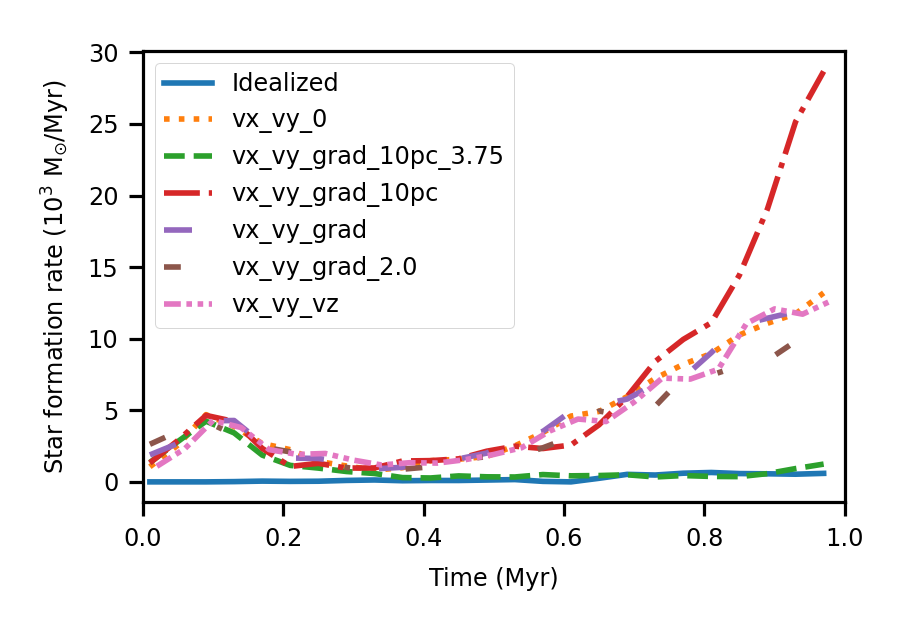}
    \caption{Star formation rate as a function of time for all our models. Most simulations have too high a star formation rate; only the simulation with the largest background of lower-density gas, and the idealized simulation without any observational filament, are forming stars at a reasonable rate.
}
    \label{fig:compare_sfr}
\end{figure}

\begin{figure}
\centering    
\includegraphics[width=\columnwidth]{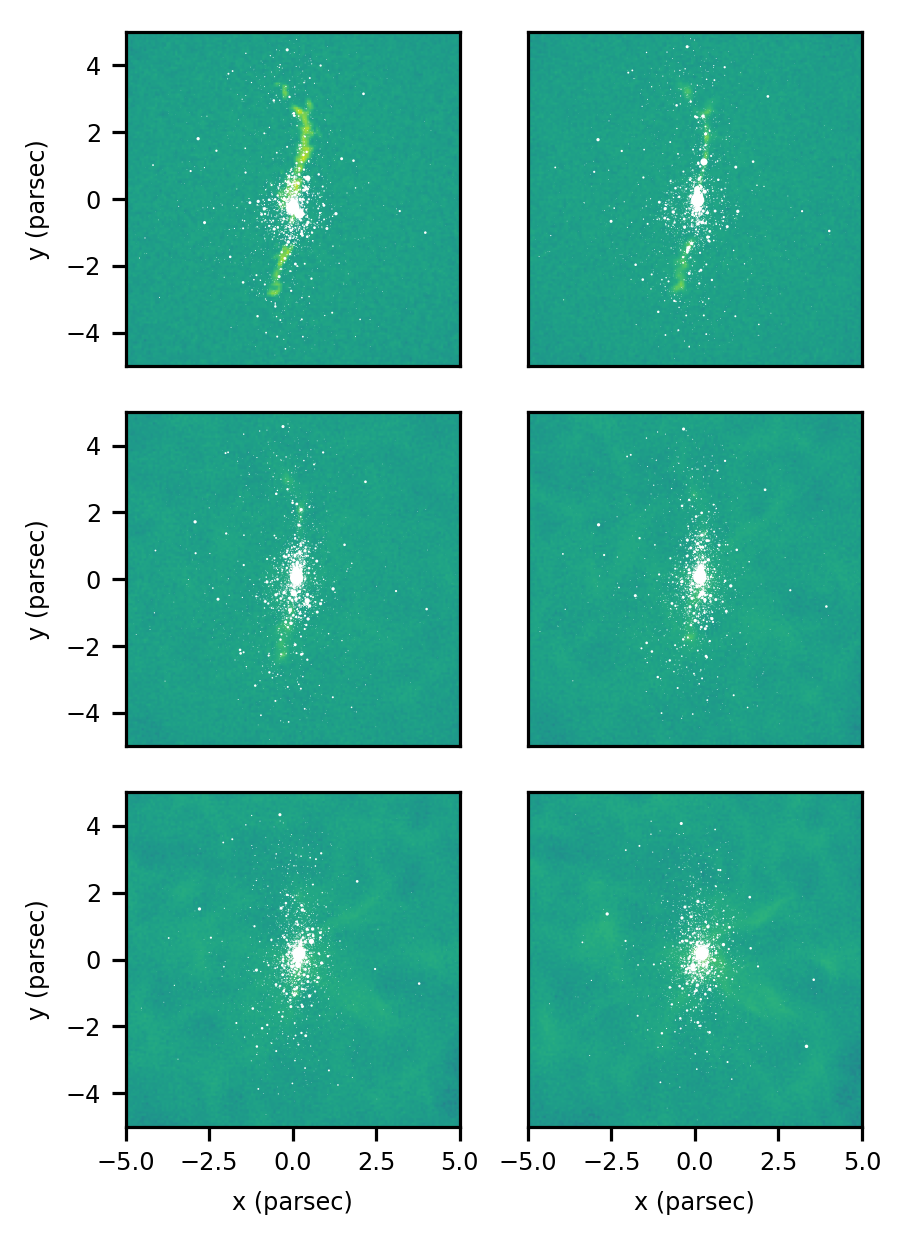}
    \caption{Evolution of model vx\_vy\_grad\_10pc\_3.75, which has the same amount of background gas but in a sphere with twice the radius. Each panel shows the projected column density and stellar positions as in Figure \ref{fig:xy}. The timestep from the top left to bottom right is 0.2 Myr. Since the background material is less dense, it collapses more slowly and the filament remains visible for longer. 
    }
    \label{fig:evolution_vxygrad_10pc_375}
\end{figure}

\begin{figure}
    \centering
    \includegraphics[width=\columnwidth]{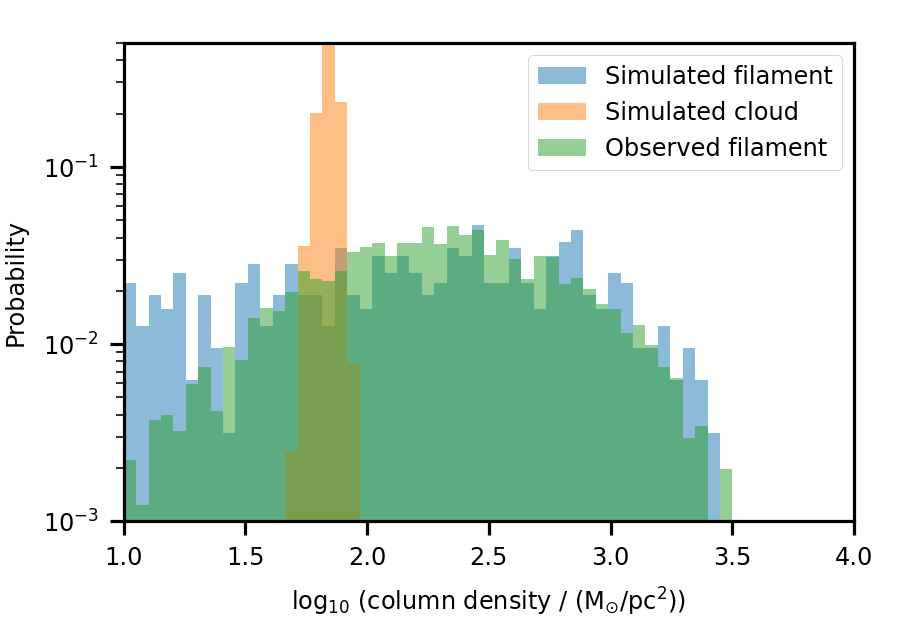}
    \caption{Probability density function of the gas column density at t=0 in the simulation shown in Figure \ref{fig:evolution_vxygrad_10pc_375}. The green histogram shows the observed dense filament, and the blue histogram shows our simulation representation, which is the same as shown in Figure \ref{fig:PDF}. 
    The orange histogram shows the background cloud. The lower density of the background component results in a lower, and more reasonable, star formation rate. 
    }
    \label{fig:PDF_best}
\end{figure}

In Figure \ref{fig:compare_sfr} we plot the star formation rate as a function of time for all the simulations described above.
The three simulations with difference choices of the gas velocity in the plane of the sky (orange dotted, purple dashed, and pink dot-dashed lines) have almost exactly the same star formation history, and therefore we conclude that the evolution of the system is relatively insensitive to any reasonable choice of velocity for the dense gas, likely due to its short free-fall time.
We can see that simply adding more gas of the same density (red line) also changes very little at early times, but allows star formation to continue for longer and at an increasing rate.
However, if we remove the filament of dense gas, and instead just follow the evolution of the background material of approximately the same high density in the 10 pc cloud (the 'idealized' run), we do not see an initial peak of star formation found in all the other runs.
This tells us that the first burst of star formation comes from the collapse of the filament.
The only simulation which does include the filament and also produces a reasonable continuous star formation rate of about 400 $M_{\odot}$ Myr$^{-1}$ is the one in which the filament is embedded in a lower-density larger sphere of material (green dashed line).
Snapshots of the evolution of this simulation are shown in Figure \ref{fig:evolution_vxygrad_10pc_375}, and its column density PDF is shown in Figure \ref{fig:PDF_best}.
The background cloud is centred at lower column density, and so much less of the gas can reach the star-forming densities even as the cloud collapses.
We conclude that the larger-scale environment of the star-forming region is critically important to its subsequent evolution. The presence of the dense filament is important to drive the first burst of star formation that we see, but the properties of the background gas have more of an effect on the star formation rate at later times. We point out that we have only tested a few parameters of a deliberately simple background of gas, and we expect that a more realistic gas distribution, preferably taken from observations but also those motivated by larger-scale simulations, would be important to include. We also note that improvements to the physics of the simulations, such as including magnetic fields, are likely to also change the evolution of the system.

\section{Evolution of stellar system and impact of gas properties}

\begin{figure}
    \includegraphics[width=\columnwidth]{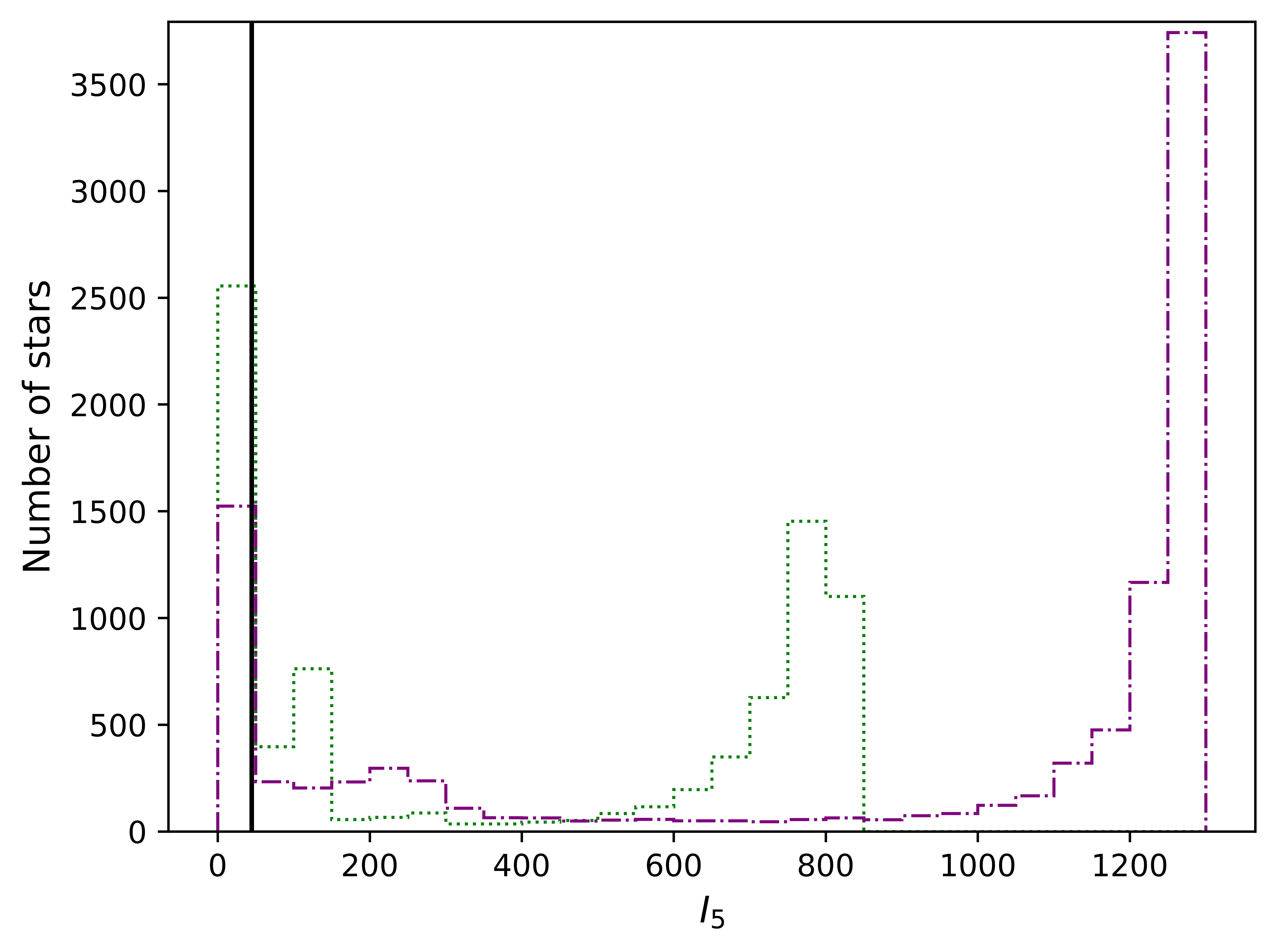}\\
   \caption{Histogram showing the distribution of index values for the population, as determined by INDICATE, at 0.1\,Myr (green dotted line) and 1.0\,Myr (purple dot-dashed line).
  The solid black lines represent the significance threshold (see text for details).}  \label{Fig_indicate_results} 
\end{figure}

Ultimately, the goal of this work is to model the formation of stellar clusters as they emerge from their natal gas clouds.
We know that open clusters are gas-free and more or less spherical, virialized systems. The structure of younger systems is more complicated and clumpy in both space and time.
There is observational evidence for age spreads and gradients in embedded clusters \citep[e.g.][]{Getman2018}, and simulations, including those described above, also predict that star formation can last at least a few Myr. In fact, multiple stellar populations have been identified inside the ONC region \citep{Beccari2017}. 
That spatial and temporal clumpiness can have implications for how larger clusters are built up, for stellar mass segregation, for the binary and planetary systems, etc.
The evolution of embedded clusters will also depend on the distribution of gas within or near the stellar systems.
Therefore, we are interested in characterizing the stellar properties in our simulations. To this end, we use the package INDICATE\footnote{\url{https://github.com/abuckner89/INDICATE}} (INdex to Define Inherent Clustering And TEndencies; \citealt{2019A&A...622A.184B}) which 
is a 2D+ local statistical tool, to assess and quantify the degree of spatial clustering of stars in our simulation.
INDICATE assigns a unit-less index $I$ to each star $j$, defined as the actual number of neighbours $N_{\bar{r}}$, and the expected number of neighbours $N$ if the star were not clustered, within fixed radius $\bar{r}$.
Following \citet{2020A&A...636A..80B} we employ a value of $N=5$ for our analysis below, i.e. 

\begin{equation}
 \\   I_{j,5}=\frac{N_{\bar{r}}}{5}.
\end{equation}

Higher values of $I_{j,5}$ represent greater degree of spatial association, and the index is calibrated against random distributions to define the `significance threshold' $I_{sig}$ - that is - the minimum value which denotes a star is spatially clustered. 
Extensive statistical testing has shown there is no dependency between the index and cluster shape, size or stellar density, and it is robust against edge effects, outliers and sample incompleteness up to $83.3\%$ (\citealt{2019A&A...622A.184B}, \citealt{2022A&A...659A..72B}). 
As a local measure of spatial association, INDICATE can also reliably identify and quantify signatures of mass segregation in clusters \citep{2022MNRAS.510.2864B}, defined as either the concentration of (I) high mass stars together, typically at a cluster's centre or (II) low/intermediate mass stars around high mass stars.
Potentially signatures of both Type I and II can be present in a cluster but the assessment is independent i.e. if low/intermediate mass stars are in concentrations around high mass stars, this does not necessarily mean that high mass stars are also concentrated together within a cluster, and vice versa.
We applied INDICATE to the simulation shown in figure \ref{fig:evolution_vxygrad_10pc_375} in time step intervals of 0.1\,Myr to follow the spatial evolution of stars.
As expected, there is a change in the behaviour of stars as the system evolves. 
Initially stars are loosely concentrated about the centre of the cluster with a median index value of 411.8 ($I_{sig}=44.5$) for the entire population of which $69.9\%$ are categorized as spatially clustered ($I_{j,5}>I_{sig}$).
A steady increase then occurs in both the number of stars that are clustered, and the degree of association, such that by 1\,Myr the median index for the population has increased to 1206.8 ($I_{sig}=44.6$) of which $84.3\%$ are categorized as spatially clustered.
This suggests that as the region evolves, not only is a greater proportion of its population likely to become clustered, but those stars which are clustered will become more spatially concentrated than at present after 1\,Myr. This is the expected behaviour of a region which is undergoing the global collapse shown in figure \ref{fig:evolution_vxygrad_10pc_375}.

\begin{figure*}
\centering
   \includegraphics[width=0.49\textwidth]{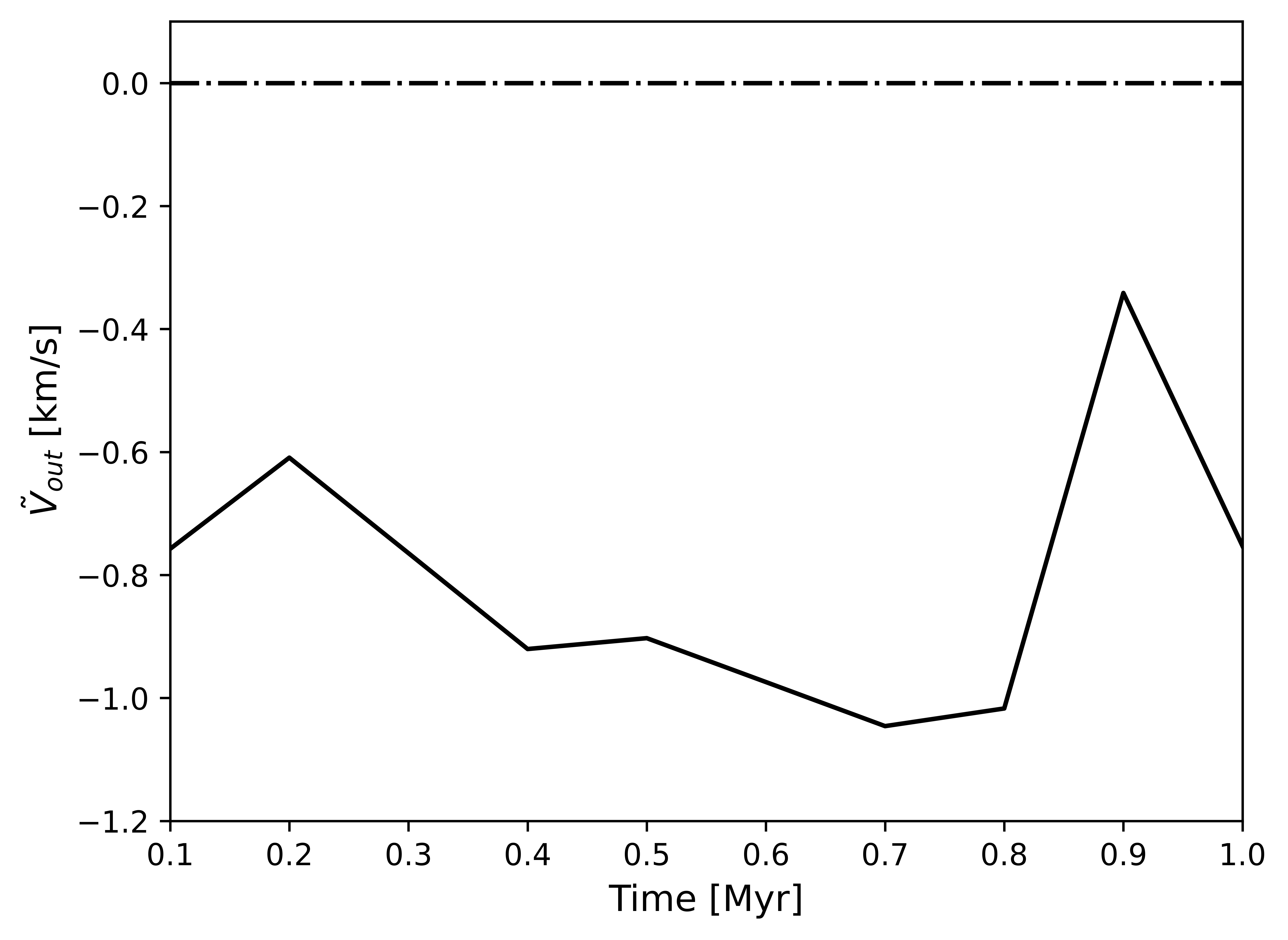}
    \includegraphics[width=0.49\textwidth]{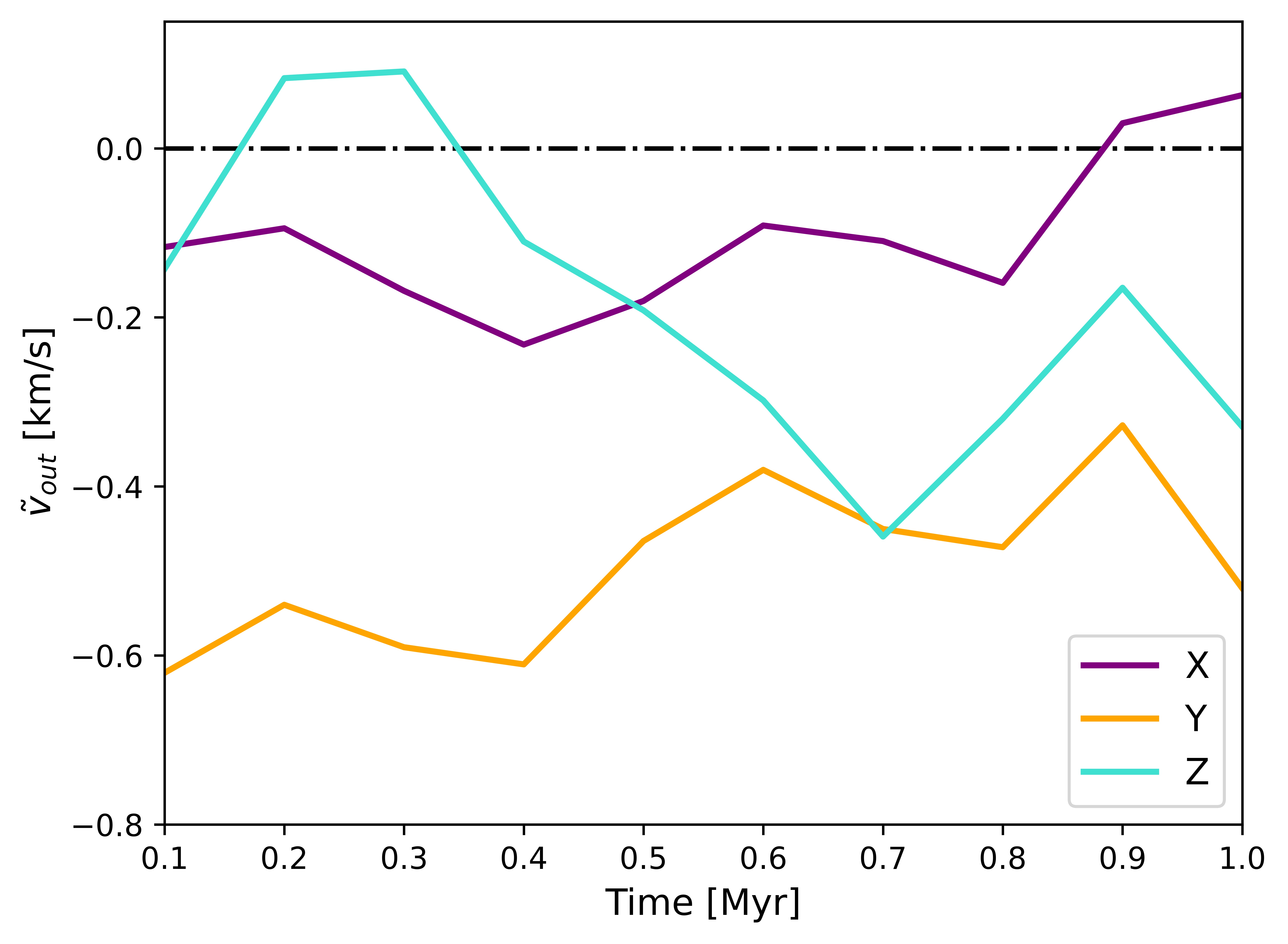}
   \caption{As a function of time after the start of the simulation,  we show (Left:) the 3D outward unit vector of velocity and (Right:) the 1D directional velocity along the X, Y and Z axes. Positive values indicate cluster expansion, negative values indicate contraction, and zero indicates a static state (marked by the horizontal black dot-dash line).}  \label{Fig_vout} 
\end{figure*}

We employed a two sample Kolmogorov-Smirnov Test (2sKST) with a strict significance boundary of $p < 0.01$ to assess the significance of the differences the found disparities in the spatial behaviours at 0.1\,Myr and 1\,Myr. 
We conclude that while there is an observed change in the spatial behaviours of stars in the simulation between 0.1\,Myr and 1\,Myr, the nature of the behaviour remains similar. Examination of the index distributions (Figure\,\ref{Fig_indicate_results}) clarifies this result.
Throughout the cluster's evolution the index distribution is bimodal with the first peak occurring at increasingly high values corresponding to the most central stars.
In contrast, the second peak occurs at a lower, approximately constant value. Values in this second peak which are greater than the significance threshold correspond to central stars at relatively larger radii than in the first peak, and below the threshold is the non-core stellar population.
This suggests that this region is contracting into a more centrally condensed configuration and will retain three spatially distinct populations: (i) a central core of stars that are very tightly clustered, encased in a (ii) `halo' of stars that are more loosely clustered, and (iii) a dispersed population at larger radii.

To determine if the stellar component of the simulation is contracting we express their velocities $v_{*}$ in terms of their 3D outward component with respect to the centre of the system follow the prescription of \citet{Kuhn2019}:

\begin{equation}
 \\ v^{*}_{out}=\vec{v}_{*}\cdot \hat{r},
\end{equation}

\begin{figure}
\centering
   \includegraphics[width=0.49\textwidth]{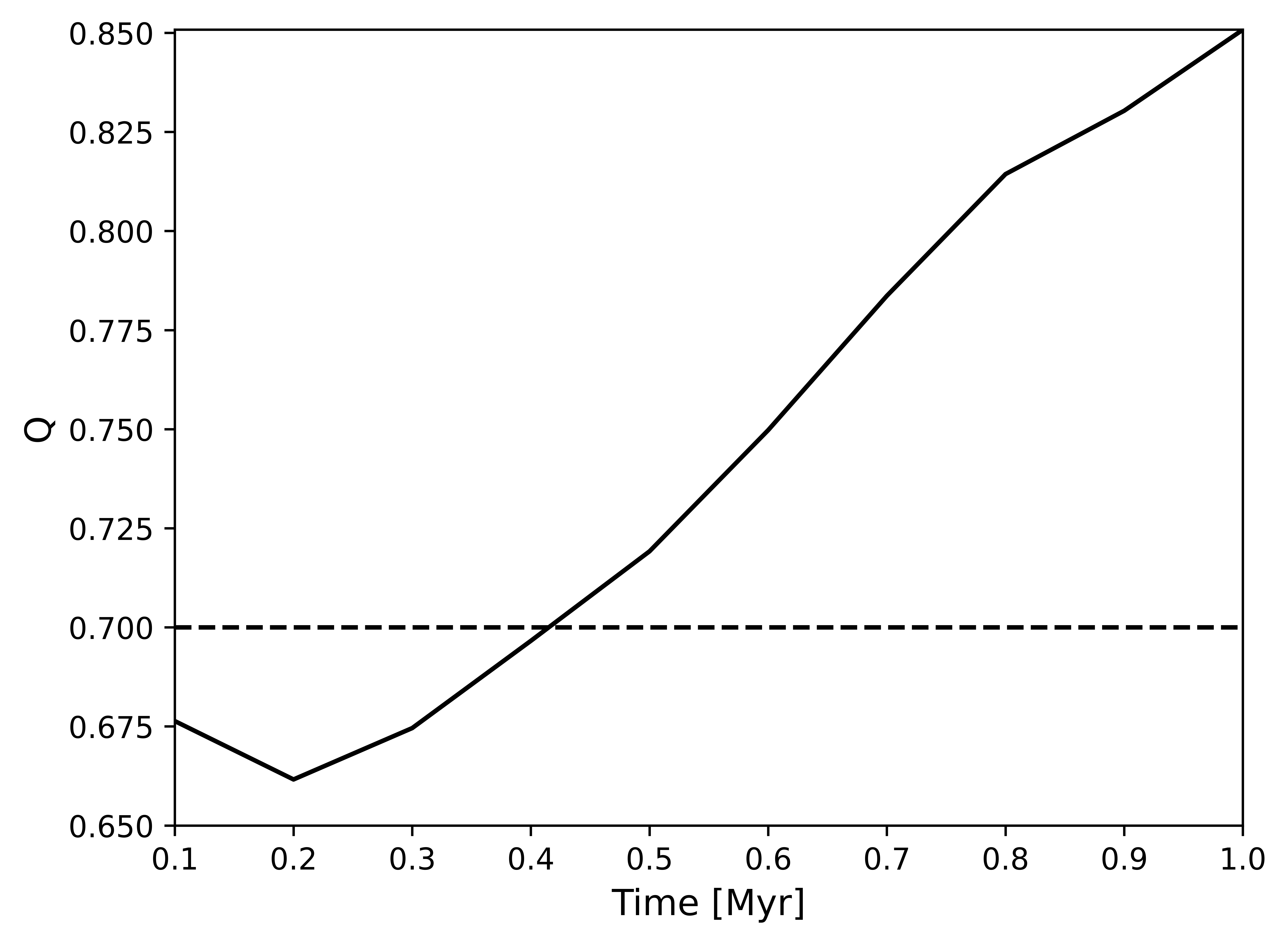}
   \caption{Plot of the $Q$ parameter (solid black line) as a function of cluster age. The dashed line represents the critical value, distinguishing between a fractal ($Q<0.7$), random ($Q\approx0.7$) and radial ($Q>0.7$) stellar distribution. }  \label{Fig_Qval} 
\end{figure}

where $\hat{r}$ is the 3D unit vector of velocity which is directed outward from the system centre. The median of these velocities $V_{out}$ indicates whether the cluster is expanding ($>0$ km s$^{-1}$) or contracting ($<0$ km s$^{-1}$).
Figure\,\ref{Fig_vout} shows $V_{out}$ and the mean 1D directional velocities as a function of cluster age in the top and bottom panels respectively.
Clearly the simulation is undergoing overall contraction at all time steps, but the 1D velocities reveal that this is occurring non-uniformly at different rates along each axis and periodically expanding in one direction.

We can also characterize the spatial distribution in the cluster by calculating the 3-dimensional Q parameter \citep{2009MNRAS.400.1427C} at each time step (Figure\,\ref{Fig_Qval}) (where a value of $Q<0.7$ indicates a fractal distribution, $Q>0.7$ radial, and $Q\approx0.7$ randomised).
We confirm that the initial distribution of stars in the simulation is a fractal distribution ( $D=2.0$), but by 0.5\,Myrs a radial distribution has been achieved that progressively strengthens through to the end point of the simulation(at which time the fractal dimension has reduced to $D=1.6$).

In particular, the high mass population (HM) is spatially evolving into an increasingly strong central concentration distribution.  We find signatures of Type I mass segregation in every snapshot with the majority of high-mass members concentrated together above spatial randomness at the simulation centre.
The degree of the segregation increases as the cluster evolves, both in fraction of stars, from $56.1\%$ at 0.1\,Myr, increasing to $79.8\%$ by 1\,Myr; and clustering strength of segregated stars, from $\tilde{I}_{5} = 84.8$ ($I_{sig}=44.3$) at 0.1\,Myr, increasing to $124.8$ ($I_{sig}=44.1$) by 1\,Myr. There is also a steady decrease observed in $\frac{r_{max}^{HM}}{r_{max}}$ from 0.31 (0.1\,Myrs) to 0.29 (1.0\,Myrs).
High mass stars are not found to be in stronger stellar concentrations in the general population than their lower mass counterparts at any point in the evolution (no signatures of Type II mass segregation are present).

\begin{figure}
\centering
   \includegraphics[width=0.49\textwidth]{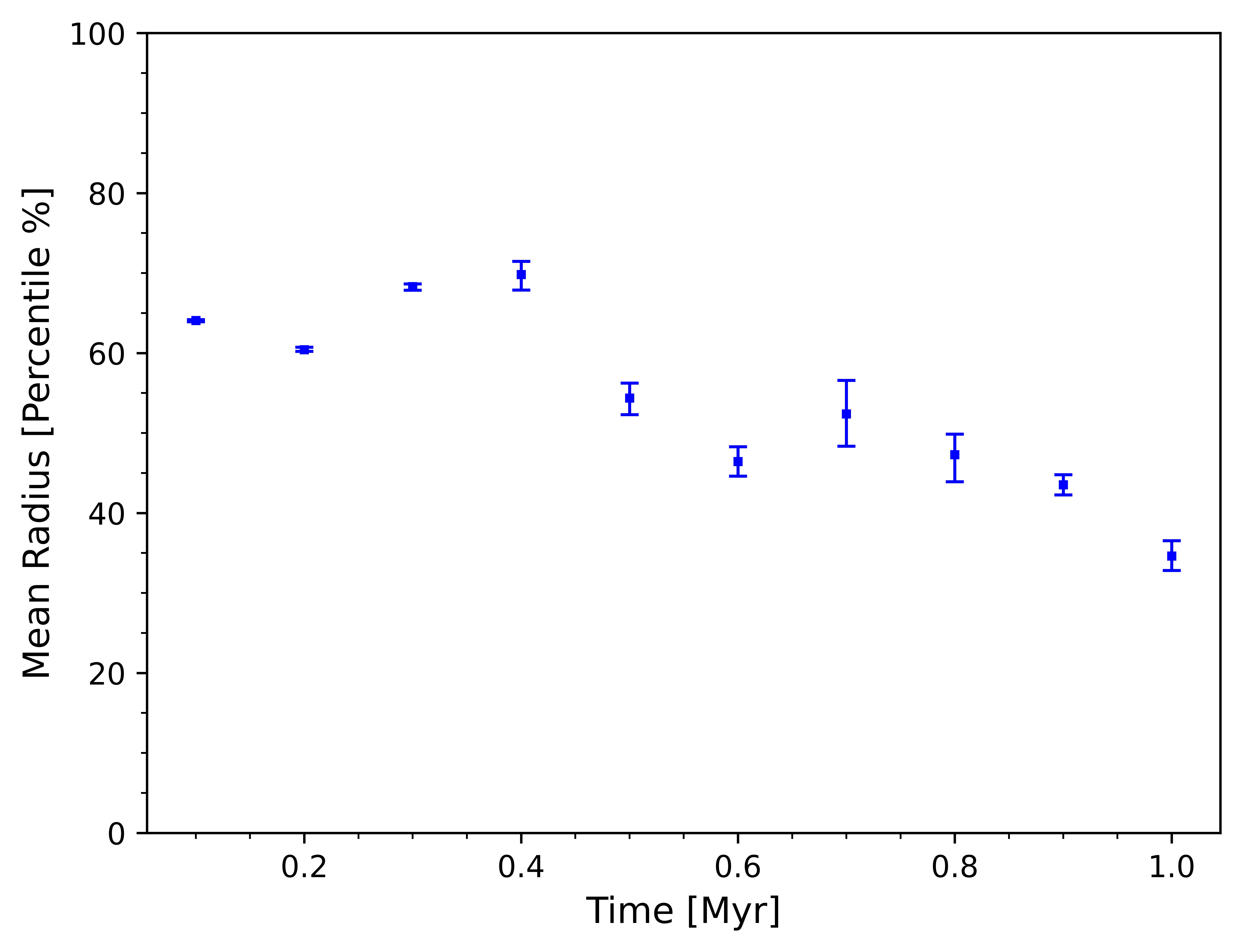}
   \caption{Mean radius from the cluster centre at which star formation is occurring (expressed as a percentile w.r.t the distribution of the pre-existing stellar population) as a function of time since the start of the simulation. Smaller percentile values represent relatively smaller radii values.}  \label{Fig_indicate_newwhere} 
\end{figure}

We consider the typical spatial conditions under which new stars are formed in the cluster.
Are stars forming in regions of progressively tighter stellar concentrations as the cluster evolves or are they forming in lower density regions and migrating to ones of higher stellar concentration?
For each time step we compute the median index values of stars that formed since the last timestep with the pre-existing population in the cluster at that time.
Prior to $\sim$0.6\,Myr stars are forming in regions of lower stellar concentrations than typical of the older cluster population, and after that time the opposite is true.
Again we run a 2sKSTs to compare the indexes of newly formed and pre-existing stars at each timestep with strict significance boundaries of $p < 0.01$. 
In all snapshots, stars that formed since the previous timestep have a distinctly different distribution to that of the pre-existing (older) population.
We calculate the distance percentile at which new stars are forming - that is, the mean radius from the cluster centre where star formation is occurring expressed as a function of the radial distribution of the pre-existing population (Figure\,\ref{Fig_indicate_newwhere}).
Initially stars form toward the outer boundary of the cluster in less populous regions, then later form with increasing centrality and in more populous regions.
As stars form in the simulation when the critical gas density criteria for star formation is met, this observed behavioural change is directly attributable to the infall of gas in the system: initially stars form less centrally where the gas has a high enough density for star formation to occur, but stellar density is low.
Then as the system evolves, gas in falls towards the cluster's centre so a critical gas density is achieved at relatively smaller radii to the simulation centre, where stellar density is higher.
Similarly, early on in the simulation a band of stars in the X-Y plane has unusually high clustering tendencies (w.r.t. the rest population in the same region), which corresponds to the location the dense gas filament shown in Figure\,\ref{fig:evolution_vxygrad_10pc_375}.

\section{Summary and Discussion}

We use observations of dense star-forming gas to guide our initial conditions of star cluster formation simulations.
This work represents an improvement on our previous efforts using observed stellar properties with only assumed gas properties. 
We outlined a method to convert the observed gas information suitable for a particle-based hydrodynamics approach.
We explored options for the quantities that are not available from the observations, particularly the velocity structure of the dense gas in the plane of the sky, and some assumptions for a wider background of gas.
We find that most reasonable assumptions for the velocity of the filament result in similar overall start formation rates and evolution of the region. We note that the presence of the filament, in any form, does have a significant effect on the star formation rate and overall structure of the forming star cluster.
The wider gas environment, however, which we modelled as a spherical configuration of less dense gas, is quite important.
In our models, only the largest cloud with the lowest density produces star formation rates which are in line with observations of nearby star-forming regions. We suggest that observations should be used to constrain all the gas in a simulation (for example, using \textit{Herschel} observations of intermediate-density gas instead of an idealized cloud).

Based on our simulations we argue that the Orion region will most likely continue to form stars, at first along the dense gas filament and later as the background gas falls into the central cluster, and at the same time, the entire region will contract into a stronger radial concentration.
The INDICATE results suggest that the degree of association in Orion will increase, but that the currently observed three spatially distinct populations (a tightly clustered core, surrounded by a loosely clustered halo, and an outer dispersed population) would be retained.
Our prediction from this work is that it is reasonable to expect Orion will become a spherical, likely gas-free cluster, within a relatively short period of time ($\sim$ 1 Myr) from the present. 

In our simulations, we restricted ourselves to single choices of some physical parameters, such as the equation of state and the prescription of star formation.
We have also neglected any feedback from the stars onto the gas including stellar winds, ionizing radiation, and protostellar jets which are expected to impact the star formation rate of low-mass clouds significantly before the first supernova can explode.
Given the sensitivity of our results to the structure of the background cloud of gas, combined with the relative insensitivity of our results to the treatment of the dense gas, we suggest that the physics that is included in star cluster formation simulations may be at least as important, and possibly more so, as the initial conditions.
Nevertheless, we have also shown that the presence of the dense filament of gas affects the early star formation rate.
It is incumbent on simulators to use as much direct information from observations as possible to improve the realism of our work. In the particular context of the formation and early evolution of star clusters, the distribution and velocity structure of the natal gas at all densities are the key properties that should be determined or guided by observations. 

Simulations of astrophysical objects and processes necessarily begin, in their earliest incarnations, with simple initial conditions. A turbulent sphere of gas with a constant density has only a few free parameters and it is straightforward to interpret the results from such a simulation. However, at some point we have learned all we can from the simple set-ups and we must turn to more complicated initial conditions. That opens up a wealth of possibilities and choices of parameters. There is a very significant need for those parameters to be constrained by observations, as otherwise the parameter space is simply too large to explore, and much of it will simply not be relevant. We have shown here that it is not very difficult to go one step even further, and to directly use the observations of gas and stars to dictate initial conditions for simulations. This approach may not be applicable to all simulations or all astrophysical objects, but it provides a useful additional method of investigating the universe. 

\section*{Acknowledgements}

AS is supported by the Natural Sciences and Engineering Research Council of Canada.
SR acknowledges funding from STFC Consolidated Grant ST/R000395/1 and the European Research Council Horizon 2020 research and innovation programme (Grant No. 833925, project STAREX). AB is funded by the European Research Council H2020-EU.1.1 ICYBOB project (Grant No. 818940).
This project has received funding from the European Research Council (ERC) under the European Union's Horizon 2020 research and innovation programme (Grant agreement No. 851435).
The authors are grateful to the referee (Mike Grudić), whose comments helped improve this article.

\section*{Data Availability}

The simulations described in this paper are available via \citep{sills_alison_2022_7015720}.




\bibliographystyle{mnras}
\bibliography{Orion} 

\bsp	
\label{lastpage}
\end{document}